\documentclass[aps,preprint]{revtex4}%
\usepackage{amsfonts}
\usepackage{amsmath}
\usepackage{amssymb}
\usepackage{subfigure}
\usepackage{graphicx}%
\usepackage{color}

\begin{document}
\title{Joule-Thomson expansion of the torus-like black hole}
\author{Jing Liang$^{a,b}$}
\email{jingliang@stu.scu.edu.cn}

\author{Wei Lin$^{a}$}
\email{linweiphy@126.com}

\author{Benrong Mu$^{a,b}$}
\email{benrongmu@cdutcm.edu.cn}

\affiliation{$^{a}$ Physics Teaching and Research section, College of Medical Technology,
Chengdu University of Traditional Chinese Medicine, Chengdu, 611137,
PR China}
\affiliation{$^{b}$Center for Theoretical Physics, College of Physics, Sichuan University, Chengdu, 610064, PR China}

\begin{abstract}
In this paper, we study Joule-Thomson effects for the torus-like black hole. The Joule-Thomson coefficients, the inversion
curves and the isenthalpic curves are studied.
Furthermore, we investigate similarities and differences between the Van der Waals fluid, the torus-like black hole and the charged AdS black
holes for the expansion. The isenthalpic curves in the $T-P$ plane are obtained. Moreover, we determine the cooling-heating regions.
\end{abstract}

\maketitle
\tableofcontents{}

\bigskip{}

\section{Introduction}
The works of Bekenstein and Hawking \cite{intro-Bekenstein:1972tm,intro-Bekenstein:1973ur,intro-Bardeen:1973gs,intro-Bekenstein:1974ax,intro-Hawking:1974rv,intro-Hawking:1974sw} make it possible to study the thermodynamic properties of black holes. As a thermodynamic system, black holes have many
interesting characteristics and have many similarities with general thermodynamic systems. In the past few decades, the study of
black hole thermodynamics has made great progress. Theoretical physicists have studied entropy \cite{intro-tHooft:1984kcu,intro-Cardy:1986ie},
thermal and non-thermal radiation \cite{intro-Zhang:2005xt,intro-Zhang:2007rv,intro-Zeng:2008zze,intro-Zeng:2008zzc}, thermodynamic
phase transition \cite{intro-Setare:2004sr,intro-Cavaglia:2004jw,intro-Chen:2002tu}, and the thermodynamic properties of black
holes in holographic and geometric frameworks \cite{intro-Zeng:2013fsa,intro-Zeng:2015tfj,intro-Zeng:2014xza,intro-Cai:2017ihd,intro-Han:2010zzc,intro-Han:2012hc,intro-Ruppeiner:2007hr,intro-Ruppeiner:2008kd}.

In recent years, it has been discovered that the thermodynamic pressure and volume of black holes can be introduced into black hole
thermodynamics. At this time, the pressure is defined as a dynamic variable by the cosmological constant. In fact, the dynamic
cosmological constant is a concept that has been hypothesized for some time \cite{intro-Teitelboim:1985dp,intro-Brown:1988kg}. Later, cosmological constants were introduced
to reflect the pressure in the space-time of black holes \cite{intro-Caldarelli:1999xj,intro-Padmanabhan:2002sha}
\begin{equation}
P=-\frac{\varLambda}{8\pi}.
\end{equation}
According to the laws of thermodynamics, pressure must have a
thermodynamic conjugate. This thermodynamic conjugate is the thermodynamic volume of the black hole \cite{intro-Dolan:2010ha,intro-Cvetic:2010jb}, i.e.,
\begin{equation}
V=\left(\frac{\partial M}{\partial P}\right)_{S,Q}.
\end{equation}
When combined with the $PV$ term \cite{intro-Dolan:2011xt} to extend thermodynamics, the mass of a black hole does not correspond to
its internal energy, but to its enthalpy \cite{intro-Kastor:2009wy}. Since the physical meaning of mass changes from internal energy to enthalpy, the
difference in thermodynamic phenomena depends on whether the $PV$ term is included. In particular, various studies have focused on
the thermodynamic application of $PV$ term \cite{intro-Kubiznak:2012wp,intro-Liang:2021voh,intro-Johnson:2014yja,intro-Caceres:2015vsa,intro-Karch:2015rpa,intro-Mu:2019bim,intro-Chen:2020zps,intro-Mu:2020szg,intro-Kubiznak:2016qmn,intro-Hennigar:2017apu,intro-Hong:2020zcf,intro-Liang:2020hjz,intro-Wei:2017vqs,intro-Gregory:2017sor,intro-Liang:2020uul}.

After taking cosmological constants into black hole thermodynamics, black holes can be further studied in extended phase space,
such as the first and second laws of thermodynamics, weak cosmic censorship conjecture, the heat engine, and Joule-Thomson expansion.
In \cite{intro-Okcu:2016tgt}, the Joule-Thomson expansion of black holes was first studied. In classical thermodynamics,
Joule-Thomson expansion is an enthalpy process, which refers to the process of gas expansion from high pressure to low
pressure through a porous plug. Subsequently, this creative work has been extended to various black holes, such as typical RN-AdS
black holes \cite{intro-Ghaffarnejad:2018exz}, d-dimensional charged AdS black holes \cite{intro-Mo:2018rgq}, $f(R)$ gravitational
charged AdS black holes \cite{intro-Chabab:2018zix}, with a global single AdS black hole of polarons \cite{intro-Rizwan:2018mpy},
conventional (Badeen)-AdS black hole \cite{intro-Pu:2019bxf}, charged AdS black hole in rainbow gravity \cite{intro-Yekta:2019wmt},
Hayward-AdS black hole \cite{intro-Guo:2019gkr}, AdS black hole with momentum relaxation \cite{intro-Cisterna:2018jqg}, Holographic
superfluid \cite{intro-DAlmeida:2018ldi}, Lovelock gravity \cite{intro-Mo:2018qkt}, Gauss-Bonnet black hole \cite{intro-Lan:2018nnp}.
The inversion curves of the heating-cooling zone separation on the plane of isenthalpy curves with different parameters are given.
The results of these papers show that the inversion curves of different black hole systems are similar. In classical thermodynamics,
Joule-Thomson expansion is an enthalpy process, which refers to the process of gas expansion from high pressure to low pressure
through a porous plug.
In this paper, we study the Joule-Thomson expansion in the torus-like black hole. Unlike previous studies, the topology of this
space-time is $S\times S\times M^{2}$. We intend to explore whether this topology will affect the Joule-Thomson expansion.

The rest of this paper is organized as follows. In Section \ref{sec:TL}, we briefly review the  thermodynamic of the torus-like black hole. In
Section \ref{sec:JT}, we first review the Joule-Thompson expansion of Van der Waals gas, and then study the Joule-Thompson
expansion of a torus-like black hole. Finally, we discuss the results in Section \ref{sec:con}.
\section{A brief review on torus-like black holes}
\label{sec:TL}
The metric of the torus-like black hole in four dimensional space is \cite{TL-Huang:1995zb, TL-Han:2019kjr}
\begin{equation}
ds^{2}=-G(r)dt^{2}+G^{-1}(r)dr^{2}+r^{2}\left(d\theta^{2}+d\psi^{2}\right),
\end{equation}
where
\begin{equation}
G(r)=-\frac{2M}{\pi r}+\frac{4Q^{2}}{\pi r^{2}}-\frac{\varLambda r^{2}}{3}.
\end{equation}
In the above equation, $M$ is the mass, $Q$ is the charge and $\varLambda$ is the cosmological constant. Usually, the cosmological
constant is considered as a thermodynamic variable. After the thermodynamic pressure of the black hole is introduced into the laws
of thermodynamics, the cosmological constant is treated as the pressure
\begin{equation}
P=-\frac{\Lambda}{8\pi}=\frac{3}{8\text{\ensuremath{\pi}}l^{2}}.
\label{eqn:P1}
\end{equation}
And the conjugate quantity of the cosmological constant is the thermodynamic volume $V$.
The mass of the black hole is given by
\begin{equation}
M=\frac{4\pi^{2}Pr_{+}^{4}+6Q^{2}}{3r_{+}},
\label{eqn:M1}
\end{equation}
where $r_+$ is the event horizon, and can obtain as the largest root of $f(r)=0$.
The entropy of the black hole is
\begin{equation}
S=\pi^{2}r_{+}^{2}.
\end{equation}
When the mass $M$ of the black hole is considered as a function of entropy $S$, charge $Q$ and pressure $P$, the mass $M$ is reformulated as
\begin{equation}
M=\frac{4PS^{2}}{3\pi\sqrt{S}}+\frac{2\pi Q^{2}}{\sqrt{S}}.
\end{equation}
The first law of thermodynamics can be expressed as
\begin{equation}
dM=TdS+\varphi dQ+VdP,
\end{equation}
and thermodynamic variables defined above is obtained as follows
\begin{equation}
T=\left(\frac{\partial M}{\partial S}\right)_{P,Q}=\frac{2\pi^{2}Pr_{+}^{4}-Q^{2}}{\pi^{2}r_{+}^{3}},
\label{eqn:T1}
\end{equation}
\begin{equation}
\varphi=\left(\frac{\partial M}{\partial Q}\right)_{S,P}=\frac{4Q}{r_{+}},
\label{eqn:phi}
\end{equation}
\begin{equation}
V=\left(\frac{\partial M}{\partial P}\right)_{S,Q}=\frac{4\pi^{2}r_{+}^{3}}{3}.
\label{eqn:V1}
\end{equation}
From Eqs. $\left(\ref{eqn:T1}\right)$ and $\left(\ref{eqn:V1}\right)$, the equation of state $P = P(V,T)$ for a torus-like black hole is
\begin{equation}
\begin{aligned}
&P=\frac{Q^{2}+\pi^{2}r_+^{3}T}{2\pi^{2}r_+^{4}},\\
&r_{+}=\left(\frac{3V}{4\pi^{2}}\right)^{\frac{1}{3}}.\\
\label{eqn:PTr}
\end{aligned}
\end{equation}

\section{Joule-Thomson expansion}
\label{sec:JT}
In this section, we review the famous Joule-Thomson expansion. In the process of Joule-Thomson expansion, the
temperature changes with the pressure, and the enthalpy does not change \cite{JT-Johnston:2014tgt}. The sign of the Joule-Thomson
coefficient $\mu$ can be used to determine whether cooling or heating, which is defined as
\begin{equation}
\mu=\left(\frac{\partial T}{\partial P}\right)_{H}.
\end{equation}
The pressure change is negative because the pressure always decreases during the expansion process. This may lead to an increase or
decrease in temperature during expansion. Therefore, the sign of the Joule-Thomson coefficient $\mu$ is determined by the change in
temperature. If it is negative (positive), heating (cooling) occurs and the fluid becomes warm (cooling). In the extended phase
space, when comparing a black hole system with a Van der Waals fluid with a fixed number of particles, a regular ensemble with a
fixed charge $q $ should be considered.

The Joule-Thomson coefficient $\mu$ can also be expressed in terms of volume and heat capacity under constant pressure. According to
the first law of thermodynamics, the fundamental relationship for the constant number of particles $N$ is written as
\begin{equation}
dU=TdS-PdV.
\end{equation}
Considering the relation $H=U+PV$, the above equation is rewritten as
\begin{equation}
dH=TdS+VdP.
\end{equation}
Since the Joule-Thomson effect occurs when the thermal expansion process is irreversible and the enthalpy is constant, $dH=0$ can be
obtained. Then
\begin{equation}
0=T\left(\frac{\partial S}{\partial P}\right)_{H}+V.
\label{eqn:JT dH0}
\end{equation}
Since entropy is a state function, $dS$ is expressed as
\begin{equation}
dS=\left(\frac{\partial S}{\partial P}\right)_{T}dP+\left(\frac{\partial S}{\partial T}\right)_{P}dT,
\end{equation}
which can be rewritten as
\begin{equation}
\left(\frac{\partial S}{\partial P}\right)_{H}=\left(\frac{\partial S}{\partial P}\right)_{T}+\left(\frac{\partial S}{\partial T}\right)_{P}\left(\frac{\partial T}{\partial P}\right)_{H}.
\end{equation}
Substitute the above expression into Eq. $\left(\ref{eqn:JT dH0}\right)$ yields
\begin{equation}
0=T\left[\left(\frac{\partial S}{\partial P}\right)_{T}+\left(\frac{\partial S}{\partial T}\right)_{P}\left(\frac{\partial T}{\partial P}\right)_{H}\right]+V.
\end{equation}
Considering the Maxwell relation $\left(\frac{\partial S}{\partial P}\right)_{T}=-\left(\frac{\partial V}{\partial T}\right)_{P}$ and
$C_{P}=T\left(\frac{\partial S}{\partial T}\right)_{P}$, the above equation is modified as
\begin{equation}
0=-T\left(\frac{\partial V}{\partial T}\right)_{P}+C_{P}\left(\frac{\partial T}{\partial P}\right)_{H}+V.
\end{equation}
From the above equation, the Joule-Thomson coefficient is obtained as follows \cite{intro-Okcu:2016tgt}
\begin{equation}
\mu=\left(\frac{\partial T}{\partial P}\right)_{H}=\frac{1}{C_{P}}\left[T\left(\frac{\partial V}{\partial T}\right)_{P}-V\right].
\end{equation}
When $\mu=0$, one obtains
\begin{equation}
T_{i}=V\left(\frac{\partial T}{\partial V}\right)_{P},
\end{equation}
which is the inversion temperature.
\subsection{Van der Waals fluids}
\label{sec:JT F}
\begin{figure}[htb]
\centering
\includegraphics[scale=0.8]{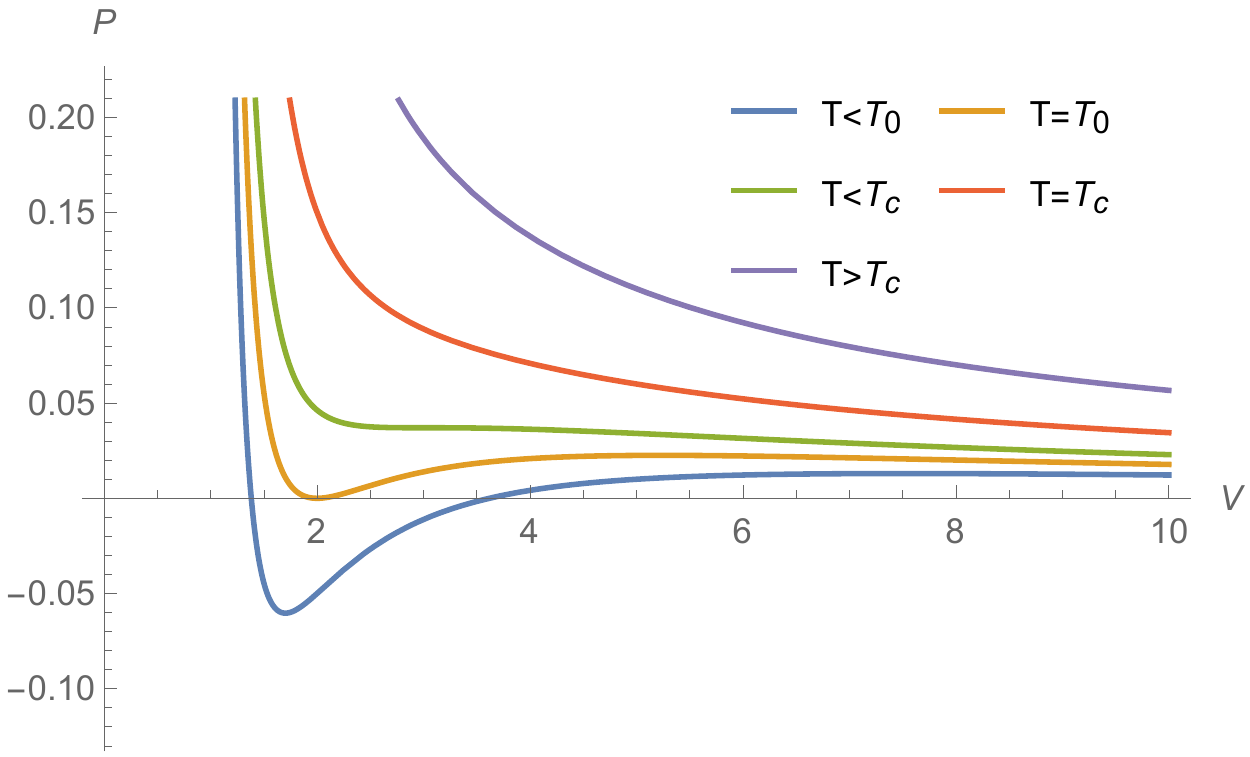}
\caption{P-V diagram of Van der Waals fluid. The temperature of the isotherm decreases sequentially from top to bottom. The constants a and b in Eq. $\left(\ref{eqn:F P}\right)$ are set to 1.}
\label{fig:F PV}
\end{figure}

Van der Waals equation is a generalized form of the ideal gas equation, which usually describes the gas-liquid phase transition
behavior of real fluids. In the Van der Waals equation, the size of the gas molecules and the interaction between them are
considered, but these are ignored in the ideal gas. The Van der Waals equation is given by
\begin{equation}
P=\frac{k_{B}T}{v-b}-\frac{a}{v^{2}},
\label{eqn:F P}
\end{equation}
where $v=\frac{V}{N}$, $P$, $T$ and $k_{B}$ are the specific volume, pressure, temperature and Boltzmann constant, respectively.
The constant $b>0$ takes into account the nonzero size of molecules in a given fluid, while the constant $a>0$ is a measure of the
attractive force between molecules. In Fig. \ref{fig:F PV}, the qualitative behavior of isotherms in the $P-V$ diagram is depicted.
The critical point appears when $P = P(r_+)$ has an inflection point, i.e.,
\begin{equation}
\frac{\partial P}{\partial r_{+}}=0,\frac{\partial^{2}P}{\partial r_{+}^{2}}=0.
\end{equation}
At the critical isotherm, $T = T_c$.

\begin{figure}[htb]
\centering
\includegraphics[scale=0.7]{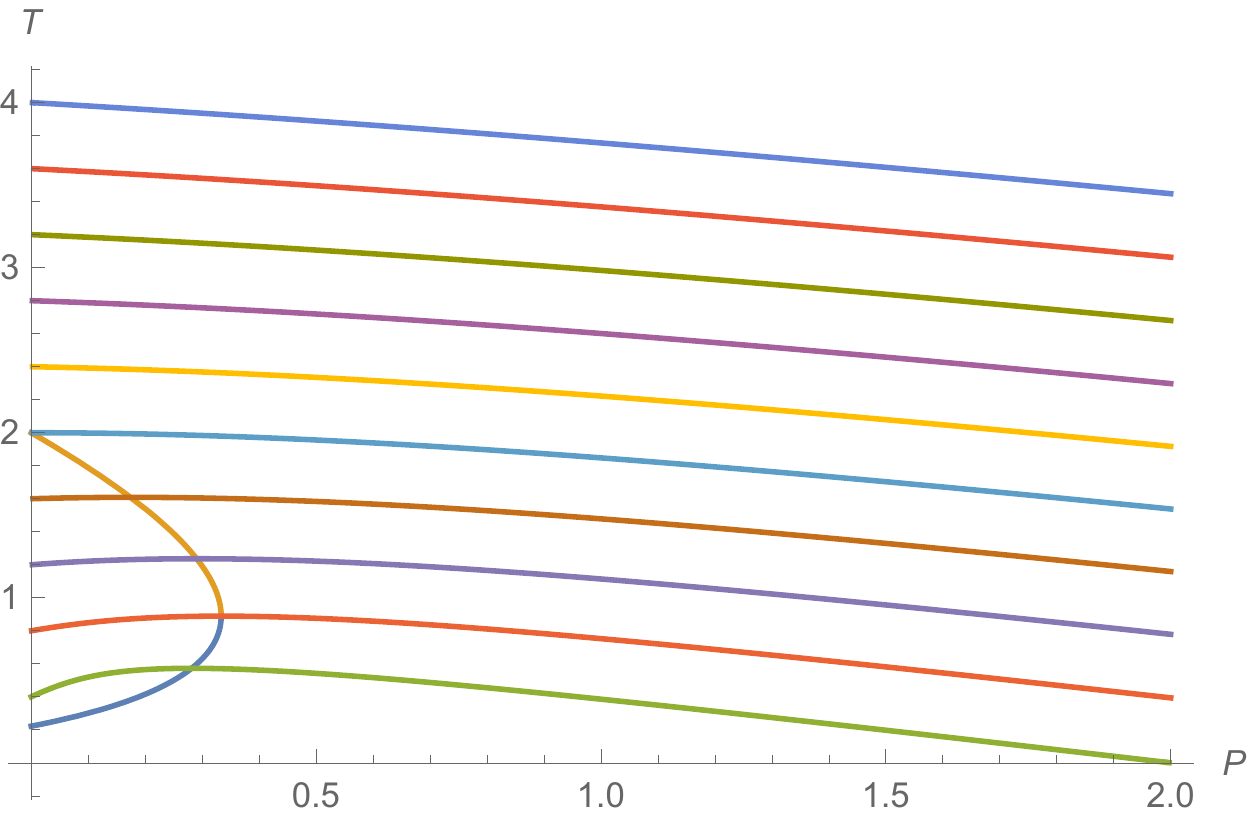}
\caption{The upper (orange line) and lower (blue line) inversion curves for the Van der Waals fluid. The colored isenthalpy curve corresponds to $H$ from 1 to 10 from bottom to top, with an interval of 1. We have set the parameters $a=b=k_B=1$ .}
\label{fig:F}
\end{figure}

The Joule-Thomson expansion of the Van der Waals system is obtained in \cite{intro-Okcu:2016tgt}. The inversion temperature of the
Van der Waals system is
\begin{equation}
T_{i}=\frac{2\left(5a-3b^{2}P_{i}\pm4\sqrt{a^{2}-3ab^{2}P_{i}}\right)}{9bk_{B}}.
\label{eqn:F TI}
\end{equation}
Based on Eqs. $\left(\ref{eqn:F P}\right)$ and $\left(\ref{eqn:F TI}\right)$, the inversion curve of the Van der Waals system is
plotted, as shown in Fig. \ref{fig:F}. The slope of the isenthalpy curve is negative in the heating zone and positive in the
cooling zone. When crossing the inversion curve, the slope of the isenthalpy curve changes.
Moreover, the minimum and maximum inversion temperatures are
\begin{equation}
T_{i}^{min}=\frac{2a}{9bk_{B}},T_{i}^{max}=\frac{2a}{bk_{B}}.
\end{equation}
The critical temperature of Van der Waals fluid is
\begin{equation}
T_{c}=\frac{8a}{27bk_{B}}.
\end{equation}
Then, the ratio of the inversion temperature to the critical temperature is
\begin{equation}
\begin{aligned}
&\frac{T_{i}^{min}}{T_{c}}=\frac{3}{4},\\
&\frac{T_{i}^{max}}{T_{c}}=\frac{27}{4}.\\
\end{aligned}
\end{equation}
\subsection{The torus-like black hole}
\label{sec:JT TL}
In this section, the Joule-Thomson expansion of a torus-like black hole is considered. As shown in Ref. \cite{intro-Kastor:2009wy},
the mass $M$ is suggested to have the physical meaning of enthalpy in extended phase space. That means that the constant enthalpy
curve is actually an equal mass curve. Therefore, we assume that the mass of the black hole does not change during the Joule-Thomson
expansion. Then
\begin{equation}
\mu=\left(\frac{\partial T}{\partial P}\right)_{M}=\frac{1}{C_{P}}\left[T\left(\frac{\partial V}{\partial T}\right)_{P}-V\right].
\label{eqn:TL mu}
\end{equation}
The equation of state is
\begin{equation}
T=\frac{6^{\frac{1}{3}}PV^{\frac{1}{3}}}{\pi^{\frac{2}{3}}}-\frac{4Q^{2}}{3V}.
\label{eqn:TL T1}
\end{equation}
Then the inversion temperature is given by
\begin{equation}
T_{i}=\frac{2^{\frac{1}{3}}P_{i}V^{\frac{1}{3}}}{(3\pi)^{\frac{2}{3}}}+\frac{4Q^{2}}{3V}=\frac{2P_{i}r_+}{3}+\frac{Q^{2}}{\pi^{2}r_+^{3}}.
\label{eqn:TL TI1}
\end{equation}
On the other hand, according to Eq. $\left(\ref{eqn:TL T1}\right)$, one obtains
\begin{equation}
T_{i}=\frac{6^{\frac{1}{3}}P_{i}V^{\frac{1}{3}}}{\pi^{\frac{2}{3}}}-\frac{4Q^{2}}{3V}=2P_{i}r_+-\frac{Q^{2}}{\pi^{2}r_+^{3}}.
\label{eqn:TL TI2}
\end{equation}
Subtracting Eq. $\left(\ref{eqn:TL TI1}\right)$ from $\left(\ref{eqn:TL TI2}\right)$ gets
\begin{equation}
2\pi^{2}r_{+}^{4}P_{i}-3Q^{2}=0.
\end{equation}
Four roots are obtained by solving this equation for $r_+$, but only one root is physically meaningful, and the other roots are
complex or negative. This positive real root is
\begin{equation}
r_{+}=\left(\frac{3}{2}\right)^{\frac{1}{4}}\frac{Q^{\frac{1}{2}}}{\pi^{\frac{1}{2}}P_{i}^{\frac{1}{4}}}.
\end{equation}
\begin{figure}[htb]
\centering
\includegraphics[scale=0.7]{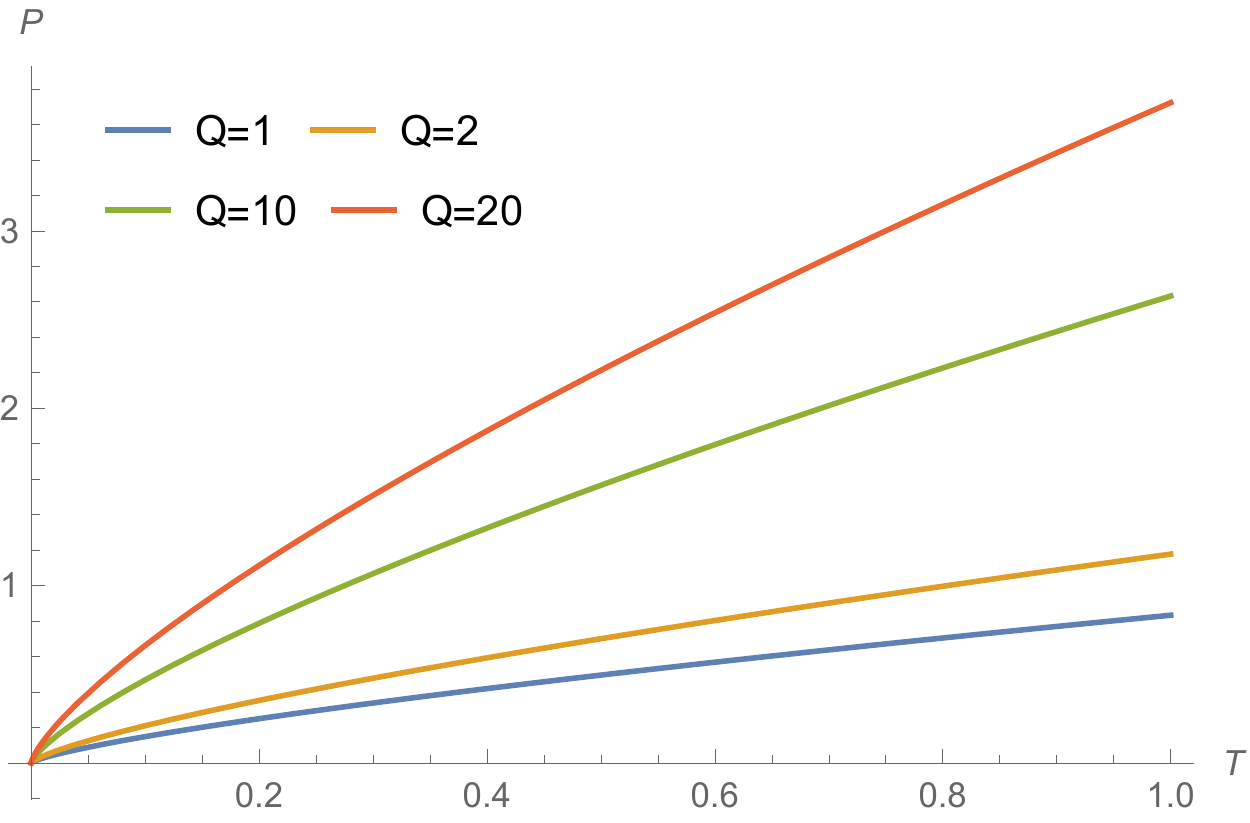}
\caption{Inversion curves for torus-like black hole.}
\label{fig:TL TP}
\end{figure}
When substituting this root into Eq. $\left(\ref{eqn:TL TI2}\right)$, the inversion temperature is written as
\begin{equation}
T_{i}=\frac{2\left(\frac{2}{3}\right)^{\frac{3}{4}}P_{i}^{\frac{3}{4}}\sqrt{Q}}{\sqrt{\pi}}.
\label{eqn:TL TI3}
\end{equation}

Inversion curves for various values of the charge $Q$ are plotted in Fig. \ref{fig:TL TP}. In the figure, there is
only the lower inversion curve. Unlike Van der Waals fluids, the expression in the square root of Eq. $\left(\ref{eqn:TL TI3}\right)$
is always positive, so inversion curves do not end at any point.

\begin{figure}[htb]
\begin{center}
\subfigure[{$Q=0.1$, $M=1.5,2.0,2.5$.}]{
\includegraphics[width=0.45\textwidth]{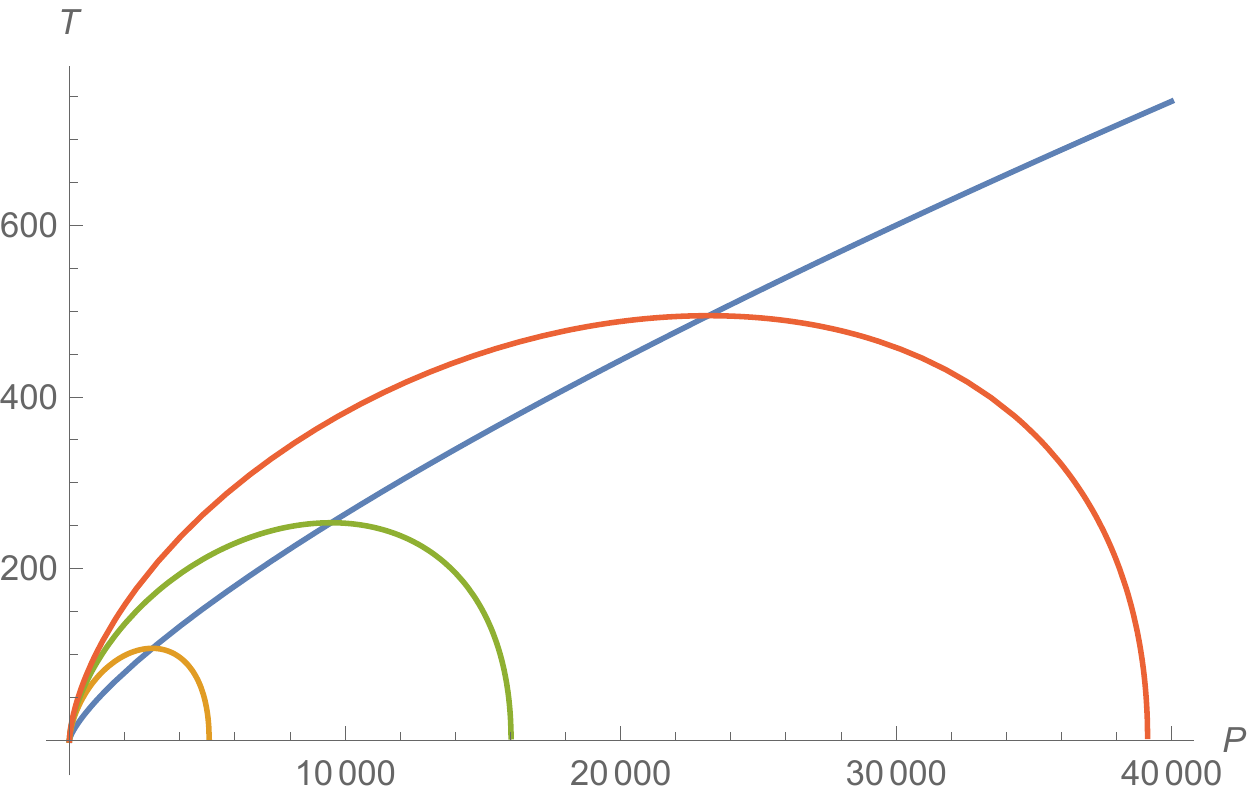}\label{fig:H01}}
\subfigure[{$Q=2$, $M=2.5,3.0,3.5,4.0$.}]{
\includegraphics[width=0.45\textwidth]{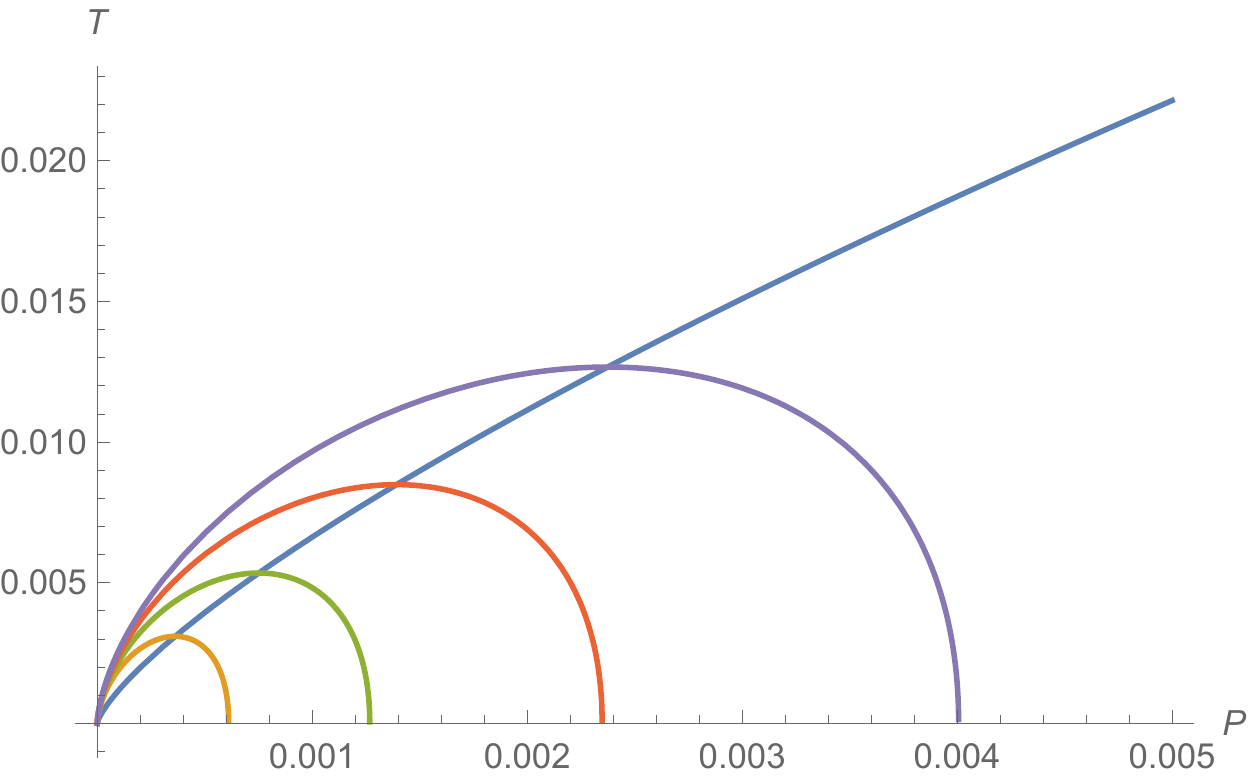}\label{fig:H2}}
\subfigure[{$Q=10$, $M=10.5,11.0,11.5,12.0$.}]{
\includegraphics[width=0.45\textwidth]{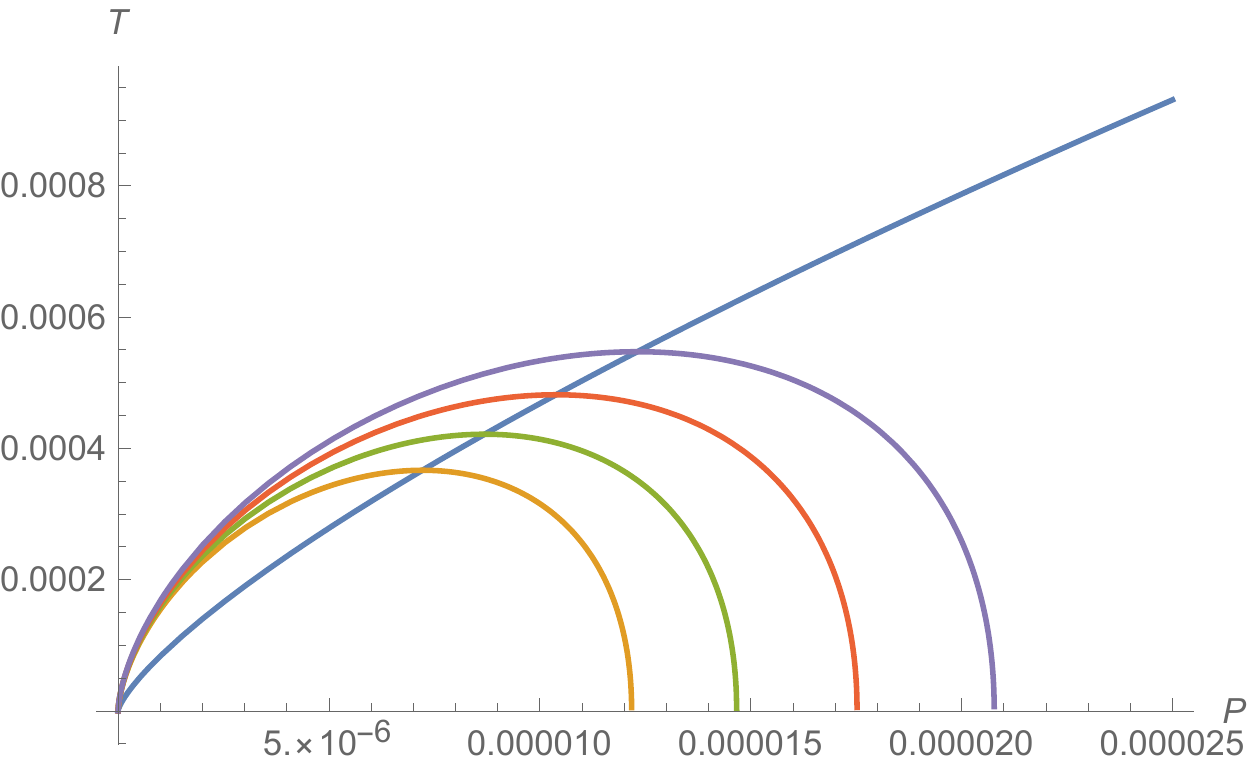}\label{fig:H10}}
\subfigure[{$Q=20$, $M=20.5,21.0,21.5,22.0$.}]{
\includegraphics[width=0.45\textwidth]{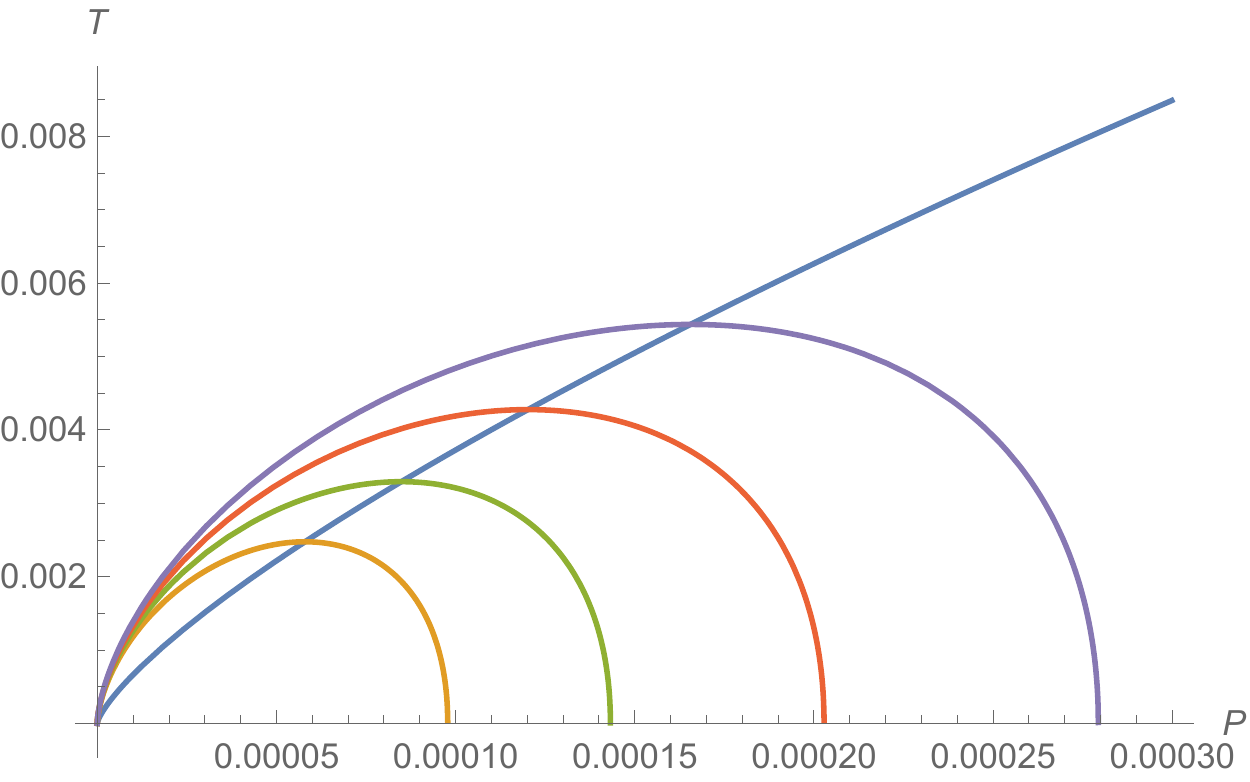}\label{fig:H20}}
\end{center}
\caption{The inversion and isenthalpic curves for the torus-like black hole. The different solid lines are isenthalpic curves with different parameters. From bottom to top, the value of $M$ corresponding to the isenthalpy curve increases. The blue lines are the inversion curves.}%
\label{fig:TLH}
\end{figure}
Joule-Thomson expansion occurs in an isenthalpic process. For a black hole, the enthalpy is the mass $M$. The isenthalpic curve can
be obtained by Eqs. $\left(\ref{eqn:M1}\right)$ and $\left(\ref{eqn:PTr}\right)$. In Fig. \ref{fig:TLH}, both the inversion curve
and the isenthalpy curve for the torus-like black hole are shown.

The intersection of the inversion curve and the isenthalpy curve coincides with the pole of the isenthalpy curve. The isenthalpy
curve has a positive slope above the inversion curve, indicating that cooling has occurred. The slope of the isenthalpy curve is
negative under the inversion curve, indicating that heating has occurred. Therefore, it can be concluded that the area above the
inversion curve is the cooling zone, and the area below is the heating zone. When the pressure in the heating zone drops, the
temperature rises. Conversely, the temperature decreases and the pressure in the cooling zone decreases. In addition, comparing
these four graphs, it can be obtained that the temperature and pressure decrease as the charge $Q$ increases.

Then, let's check whether there is a thermal phase transition of a black hole. The critical point appears when $P = P(r)$ has an inflection point,
i.e.,
\begin{equation}
\frac{\partial P}{\partial r_{+}}=\frac{\partial^{2}P}{\partial r_{+}^{2}}=0.
\end{equation}
The equation of state $P = P(r_+,T)$ for a torus-like
black hole is
\begin{equation}
P=\frac{Q^{2}+\pi^{2}r_+^{3}T}{2\pi^{2}r_+^{4}}.
\end{equation}
\begin{figure}[htb]
\begin{center}
\subfigure[{$P-V$ diagram of the torus-like black hole.}]{
\includegraphics[width=0.45\textwidth]{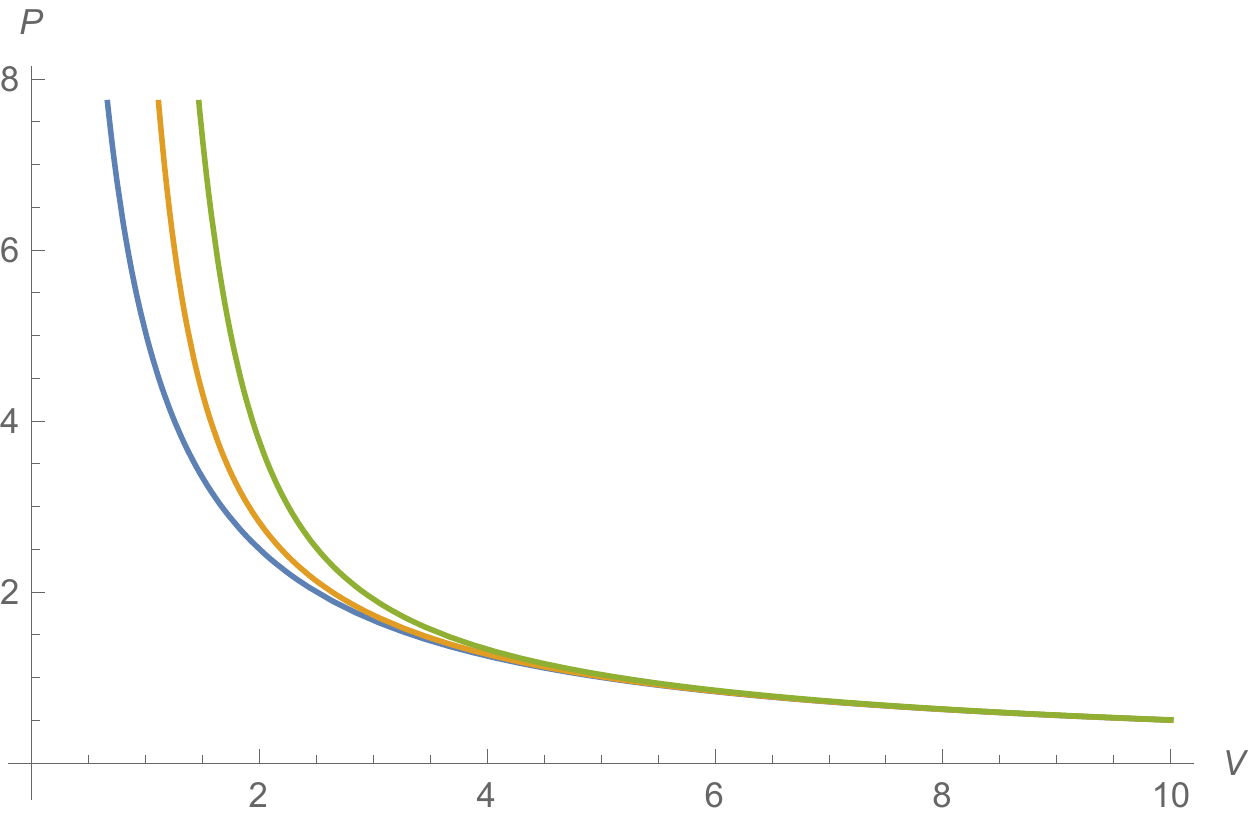}\label{fig:TLPV}}
\subfigure[{$P-V$ diagram of the charged AdS black hole.}]{
\includegraphics[width=0.45\textwidth]{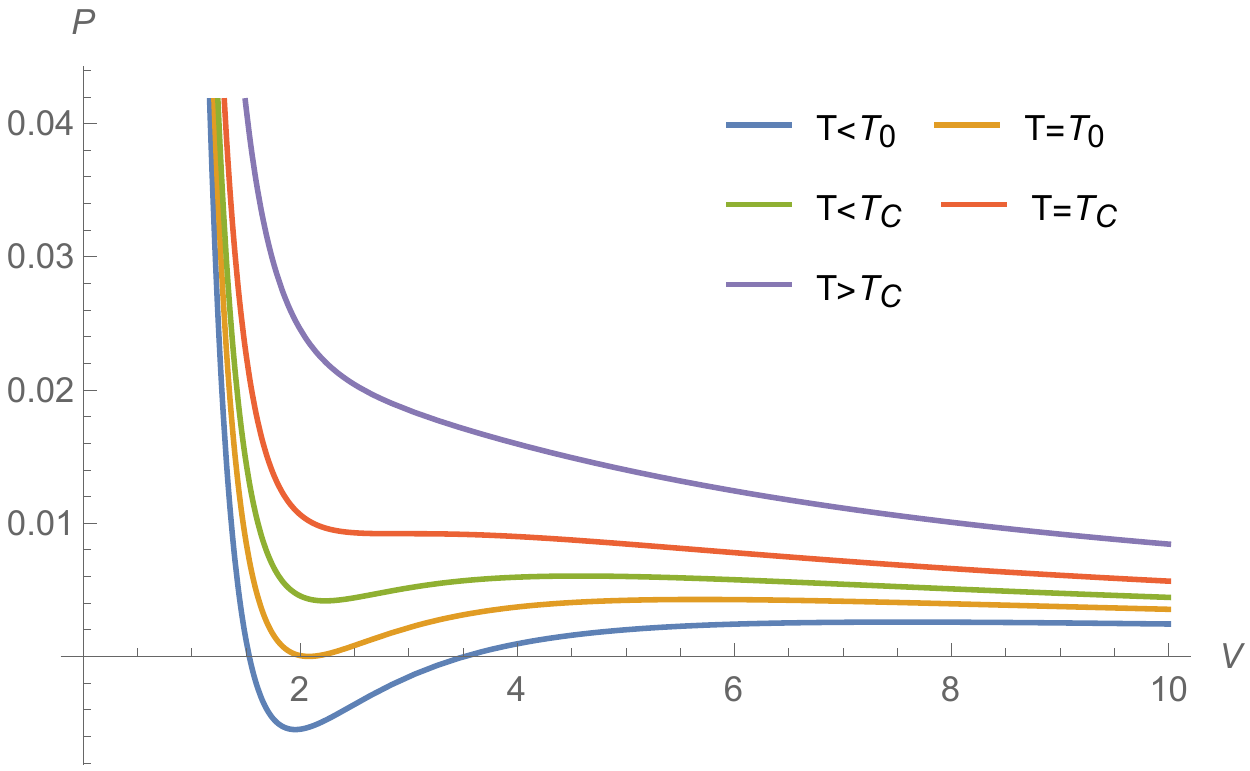}\label{fig:AdSPV}}
\end{center}
\caption{$P-V$ diagram of the torus-like black hole (left) and the charged AdS black hole (right).}
\label{fig:PV}
\end{figure}
At the critical isotherm, $T=T_c$ and $r_+=r_c$. When $\frac{\partial P}{\partial r_{+}}=0$ is satisfied, the temperature $T=-\frac{4Q^{2}}{\pi^{2}r_{+}^{3}}$.
Since the temperature is negative, there is no thermal phase change. Substituting the temperature $T$ into $\frac{\partial^{2}P}{\partial r_{+}^{2}}$ yields
\begin{equation}
\frac{\partial^{2}P}{\partial r_{+}^{2}}=\frac{6 Q^2}{\pi ^2 r_{+}^6},
\end{equation}
which can not be zero. Therefore, for the torus-like black hole, there is no thermal phase transition.
As shown in Fig. \ref{fig:TLPV}, there is no inflection point in the $P-V$ diagram of the torus-like black hole. However, the $P-V$ diagram of the charged AdS black hole has the inflection point as shown in Fig. \ref{fig:AdSPV}. The torus-black hole is always thermodynamically stable.
Therefore, for the torus-like black hole, the critical point does not exist and $T_{min}=0$. Then, the black hole becomes an extremal black hole.

\section{conclusion and discussion}
\label{sec:con}
In this paper, we took the torus-like black hole as an example to discuss in detail the influence of the topology of the space-time of
Joule-Thomson expansion. Joule-Thomson expansion is a temperature change phenomenon that describes the expansion of gas from a high
pressure section to a low pressure section through a porous plug. First, we briefly reviewed the thermodynamics of the torus-like
black hole. Then, the Joule Thomson expansion of Van der Waals fluid and the torus-like black hole were studied. The Joule-Thomson coefficient is an important
physical quantity, and its sign can be used to determine whether heating or cooling will occur. For the Van der Waals fluid, $P = P(r_+)$ has an inflection point
and the critical point exists. The minimum inversion temperature and the critical temperature were calculated. Moreover, the ratio of the minimum inversion temperature
$T_{min}$ to the critical temperature $T_c$ was calculated. For the torus-like black hole, we found that $P = P(r_+)$ doesn't have an inflection point
and the critical point doesn't exist. Furthermore, the minimum inversion temperature is zero and the black hole becomes an extremal black hole.
Since the main characteristic of Joule-Thomson expansion is that the enthalpy remains constant, we have also discussed the
isenthalpic curves of the Van der Waals fluid and black holes. In extended phase space, the mass of the black hole
should be interpreted as enthalpy. As shown in Figs. \ref{fig:F} and Fig. \ref{fig:TLH}, inversion curves and isenthalpic curves of
the Van der Waals fluid and the torus-like black hole are presented, respectively.

\begin{figure}[htb]
\centering
\includegraphics[scale=0.7]{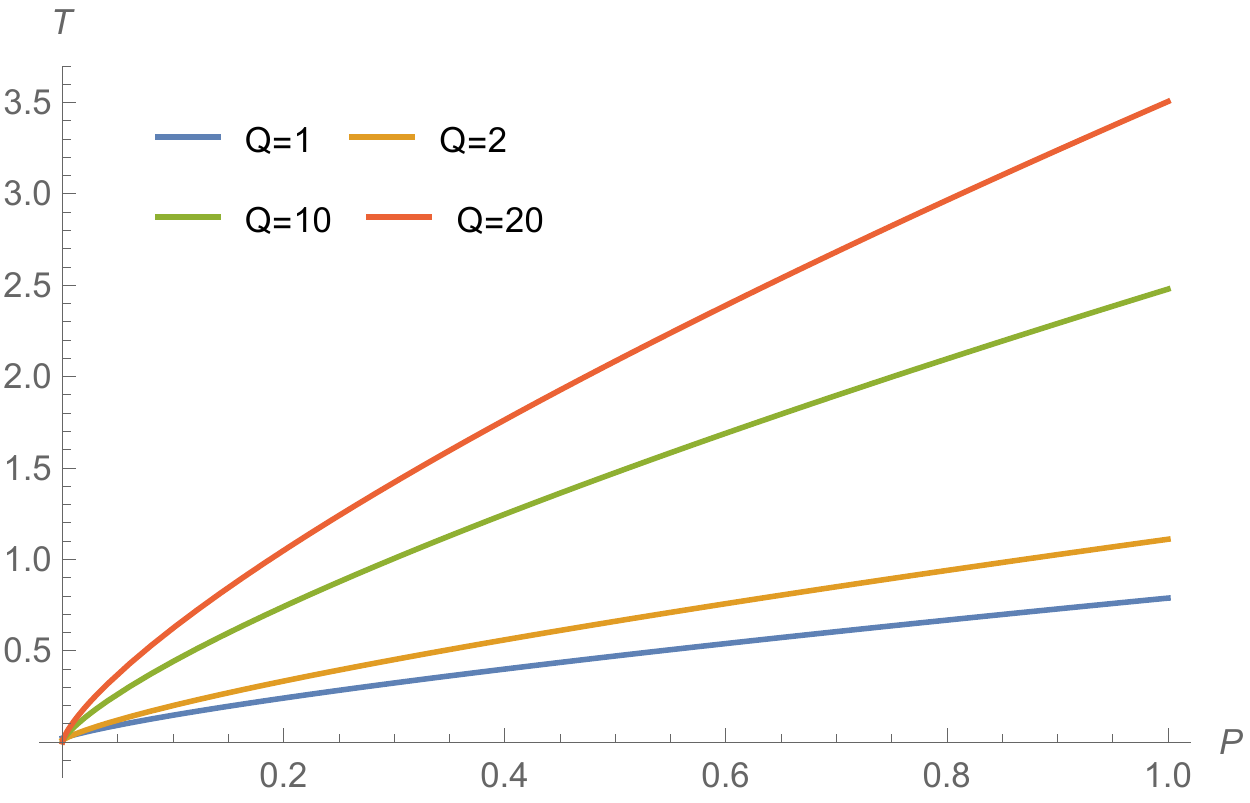}
\caption{Inversion curves for charged AdS black hole. From bottom to top, the curves correspond to $Q = 1,2,10,20$.}
\label{fig:AdSTP}
\end{figure}
The Joule-Thomson expansion of the charged AdS black hole was first studied in detail in Ref. \cite{intro-Okcu:2016tgt}.
The equation of state $P = P(r_{+},T)$ for a charged AdS black hole is
\begin{equation}
P=\frac{T}{2r_{+}}-\frac{1}{8\pi r_{+}^{2}}+\frac{Q^{2}}{8\pi r_{+}^{4}}.
\end{equation}
\begin{figure}[htb]
\begin{center}
\subfigure[{$Q=1$, $M=1.5,2.0,2.5,3.0$.}]{
\includegraphics[width=0.45\textwidth]{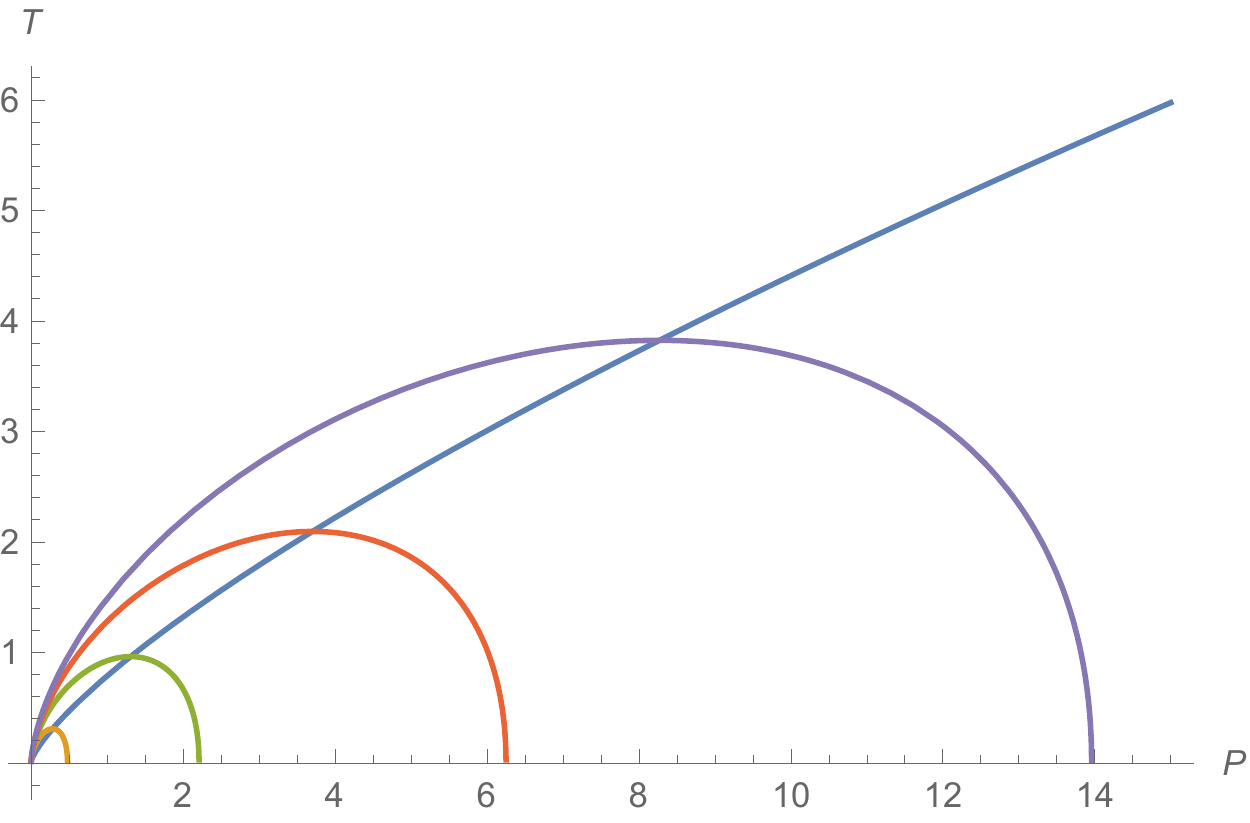}\label{fig:AdSH1}}
\subfigure[{$Q=2$, $M=2.5,3.0,3.5,4.0$.}]{
\includegraphics[width=0.45\textwidth]{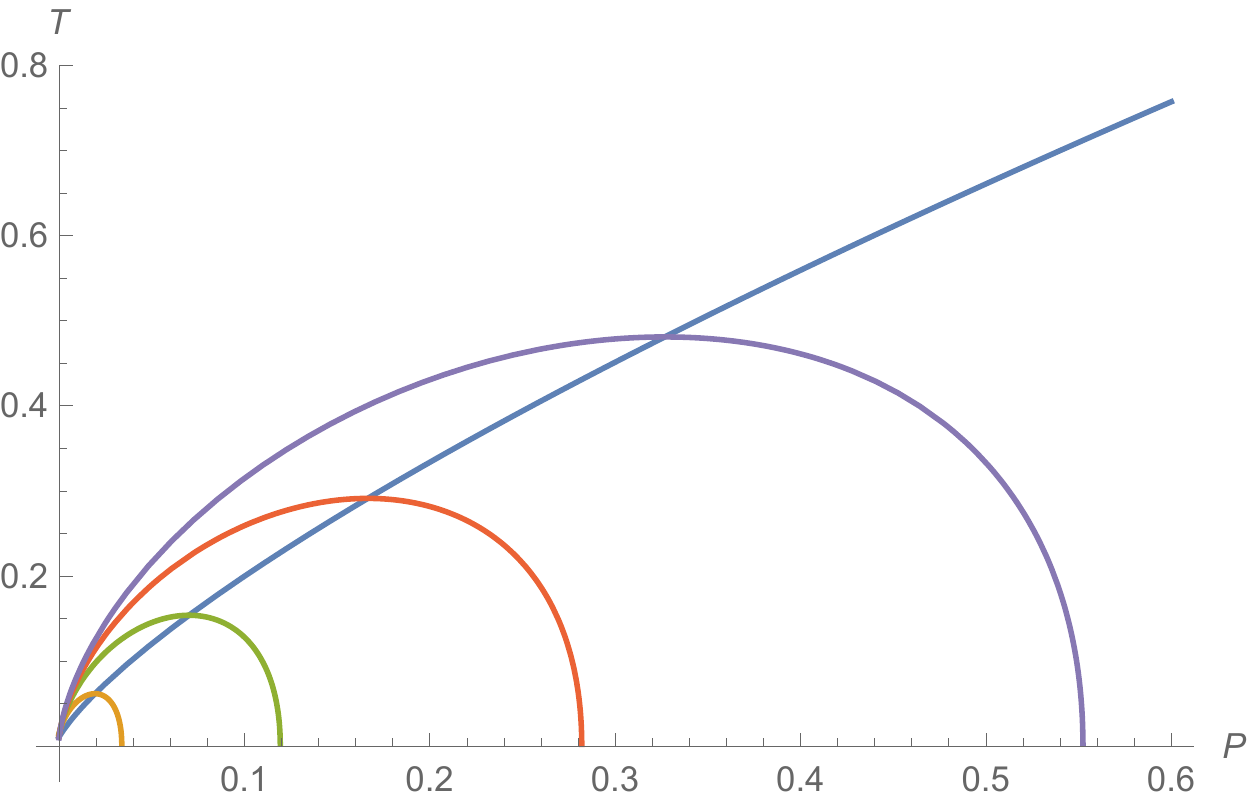}\label{fig:AdSH2}}
\subfigure[{$Q=10$, $M=10.5,11.0,11.5,12.0$.}]{
\includegraphics[width=0.45\textwidth]{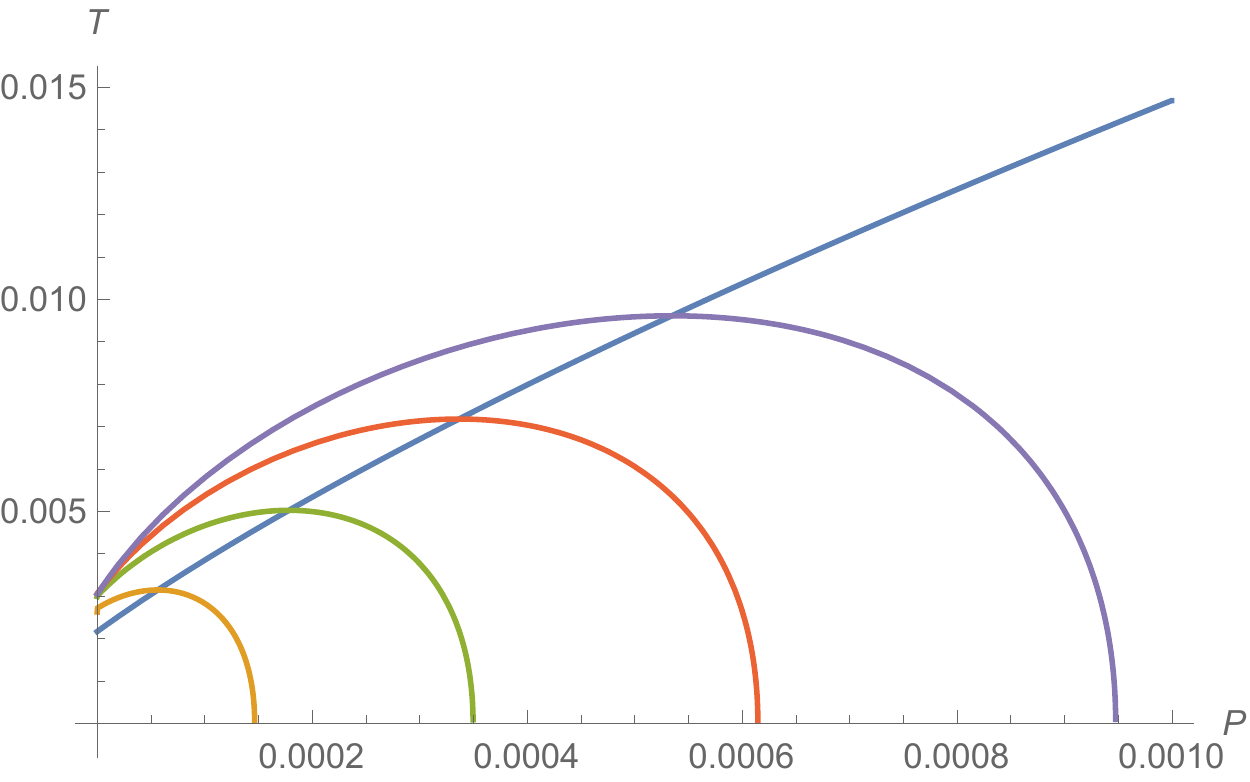}\label{fig:AdSH10}}
\subfigure[{$Q=20$, $M=20.5,21.0,21.5,22.0$.}]{
\includegraphics[width=0.45\textwidth]{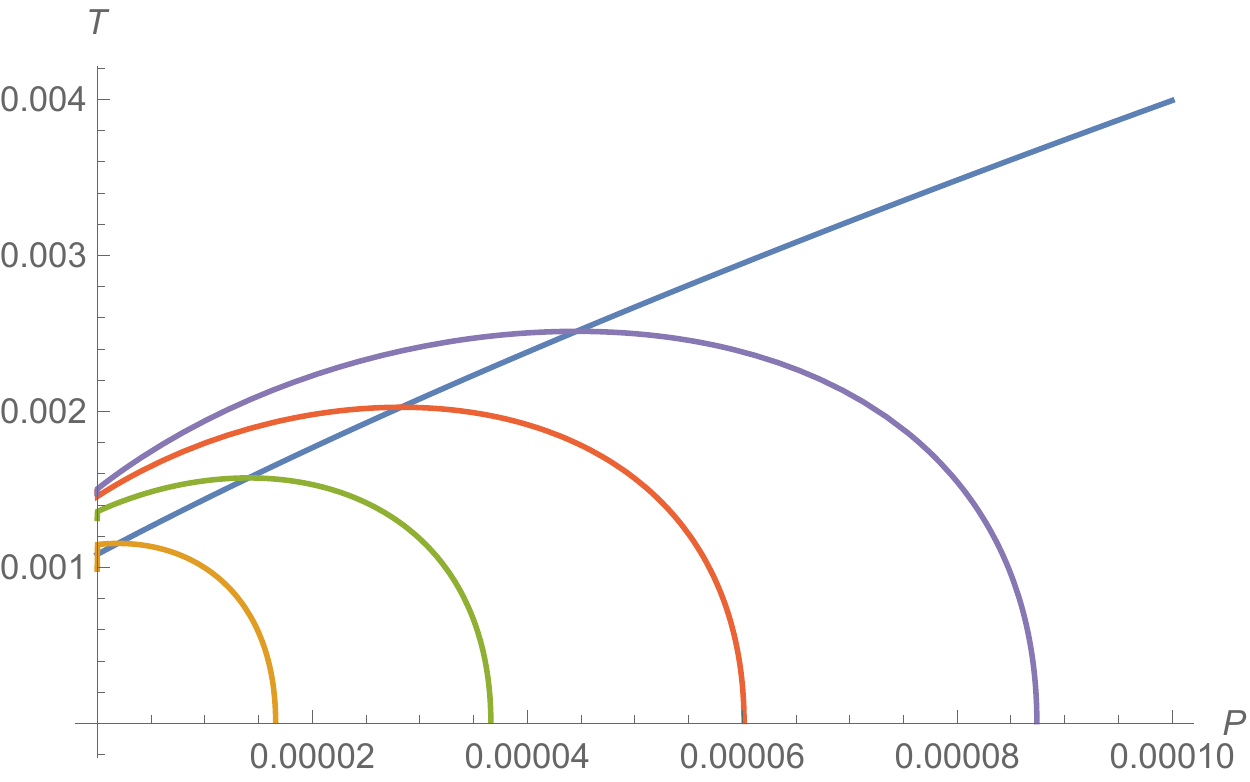}\label{fig:AdSH20}}
\end{center}
\caption{The inversion and isenthalpic curves for the charged AdS black hole. The different solid lines are isenthalpic curves with different parameters. From bottom to top, the value of $M$ corresponding to the isenthalpy curve increases. The blue lines are the inversion curves.}%
\label{fig:AdSH}
\end{figure}
The critical points are obtained from
\begin{equation}
\frac{\partial P}{\partial r_{+}}=0,\frac{\partial^{2}P}{\partial r_{+}^{2}}=0.
\end{equation}
Then, the critical temperatures of the charged AdS black hole is
\begin{equation}
T_{c}=\frac{\sqrt{6}}{18\pi Q},
\end{equation}
and the minimum inversion is
\begin{equation}
T_{i}^{min}=\frac{1}{6\sqrt{6}\pi Q}.
\end{equation}
Thus, the ratio between minimum inversion and critical temperatures is
\begin{equation}
\frac{T_{i}^{min}}{T_{c}}=\frac{1}{2}.
\end{equation}
As shown in Fig. \ref{fig:AdSTP}, the inversion curves for different charge values of the AdS black hole are plotted, and there is only a lower inversion curve.
Unlike Van der Waals fluid, the inversion curve does not end at any point.

In Table \ref{tab:ratio}, we list the existence of critical point and the ratio of the minimum inversion temperature to the critical temperature of the Van der Waals fluid, the torus-like black hole
and the charged AdS black hole.

\begin{table}[htb]
\begin{centering}
\begin{tabular}{p{1.3in}|p{0.8in}|p{0.9in}|p{1.1in}|p{0.8in}|p{0.7in}}
\hline
type  &the critical point &the critical temperature & the minimum inversion temperature & ratio & literature\tabularnewline
\hline
Van der Waals fluid & exist &$T_{c}=\frac{\sqrt{6}}{18\pi Q}$ & $T_{i}^{min}=\frac{2a}{9bk_{B}}$  & 0.75 & \cite{intro-Okcu:2016tgt} \tabularnewline
torus-like BH     & not exist  &not exist &$T_{i}^{min}=0$   & not exist & \tabularnewline
RN-AdS BH & exist &$T_{c}=\frac{\sqrt{6}}{18\pi Q},$ &$T_{i}^{min}=\frac{1}{6\sqrt{6}\pi Q}$   & 0.5 & \cite{intro-Ghaffarnejad:2018exz} \tabularnewline
\hline
\end{tabular}
\par\end{centering}
\caption{{\footnotesize{}{}{}{}The existence of critical points and the ratio of the minimum inversion temperature to the critical temperature of various black holes.}}
\label{tab:ratio}
\end{table}
The critical behavior only exists in the black hole with horizon of spherical topological \cite{intro-Kubiznak:2012wp}. For different topologies the equation of state is
\begin{equation}
P=\frac{T}{2r_{+}}-\frac{k}{8\pi r_{+}^{2}}+\frac{Q^{2}}{8\pi r_{+}^{4}},
\end{equation}
where $k=-1$ is the case of the higher-genus (hyperbolic) and $k=0$ is the case of the toroidal (planar).
For the torus-like black hole, it has the horizon of toroidal topology. For the charged AdS black hole, it has the horizon of spherical topology. As shown in Fig. \ref{fig:PV} and Table \ref{tab:ratio}, the charged AdS black hole has an inflection point for $Q\neq0$ and $T<T_c$. The behaviour of the AdS black hole is reminiscent of the Van der Waals gas. However, the torus-like black hole does not have an inflection point. Moreover, there are Joule-Thomson expansion both for the torus-like black hole and the charged AdS black hole.
Therefore, the topological structure of space-time has an impact on critical points, and thus on the critical temperature and the ratio of the minimum inversion temperature to the critical temperature. In addition, the topology of space-time
does not affect whether there is the Joule-Thomson expansion in a black hole.

\begin{acknowledgments}
We are grateful to Wei Hong, Peng Wang, Haitang Yang, Jun Tao, Deyou Chen and Xiaobo Guo for useful discussions. The authors contributed equally to this work. This work is supported in part by NSFC (Grant No. 11747171), Natural Science Foundation of Chengdu University of TCM (Grants nos. ZRYY1729 and ZRYY1921), Discipline Talent Promotion Program of /Xinglin Scholars(Grant no.QNXZ2018050) and the key fund project for Education Department of Sichuan (Grantno. 18ZA0173).
\end{acknowledgments}

\end{document}